\documentclass[preprints,article,accept,oneauthor]{Definitions/mdpi}
\firstpage{1}
\pubvolume{1}
\issuenum{1}
\articlenumber{0}
\pubyear{2023}
\copyrightyear{2023}
\datereceived{ }
\dateaccepted{ }
\datepublished{ }
\hreflink{https://doi.org/}
\pdfoutput=1

\Title{Mission Target: Exotic Multiquark Hadrons --- Sharpened Blades}
\TitleCitation{Mission Target: Exotic Multiquark Hadrons --- Sharpened Blades}

\Author{Wolfgang Lucha \orcidA}
\AuthorNames{Wolfgang Lucha}
\AuthorCitation{Lucha, W.}
\address[1]{Institute for High Energy Physics, Austrian Academy of Sciences, Nikolsdorfergasse 18, A-1050 Vienna, Austria}
\corres{Correspondence: Wolfgang.Lucha@oeaw.ac.at (W.L.)}
\abstract{Motivated by recent experimental progress in establishing the likely existence of (variants of) exotic hadrons, predicted to be formed by the strong interactions, various proposed concepts~and~ideas are compiled in an attempt to draft a coherent picture of the achievable improvement in the theoretical interpretation of exotic hadrons in terms of the underlying quantum field theory of strong interactions.}
\begin{document}

\section{Significance of Fundamental Diverseness of Ordinary Hadrons and Multiquark States}\label{sE}Within the framework of (relativistic) quantum field theories, all strong interactions are described --- at fundamental level --- by quantum chromodynamics (QCD), a renormalizable gauge theory, invariant under local transformations forming a representation of the compact non-Abelian Lie group SU(3). Two sorts of particles constitute the (basic) dynamical degrees of freedom of QCD: massless vector gauge bosons labelled gluons, transforming~(inevitably) according to the eight-dimensional adjoint representation $\mathbf{8}$ of SU(3), and spin-$\frac{1}{2}$ fermions~$q_a$, labelled quarks, each distinguished from all others by some quark flavour degree of freedom\begin{equation}a\in\{u,d,s,c,b,t(,\dots?)\}\label{qf}\end{equation}and transforming according to the three-dimensional fundamental representation $\mathbf{3}$ of~SU(3). The (few) fundamental parameters characterizing QCD are the masses $m_a$ of the quarks~$q_a$~as well as the strong coupling $g_{\rm s}$, frequently adopted in form of a strong fine-structure coupling\begin{equation}\alpha_{\rm s}\equiv\frac{g_{\rm s}^2}{4\pi}\ .\label{as}\end{equation}This designation as quantum \emph{chromo}dynamics derives from the fact that the quark and~gluon degree of freedom affected by their gauge-group transformation is referred to as their colour.

Among others, QCD features the phenomenon of colour \emph{confinement}: not the (coloured) quarks and gluons but \emph{exclusively} their colour-singlet \emph{hadron} bound states \cite{LS&G} invariant under the action of the QCD gauge group are, in form of isolated states, experimentally~observable. Closer inspection reveals that the hadron states \emph{have to be} divided into two \emph{disjoint} categories:\begin{itemize}\item\textbf{Conventional} (ordinary) hadrons include all mesons that consist of only a pair of quark and antiquark, as well as all baryons that consist of three quarks or of three~antiquarks.\item\textbf{Exotic} hadrons are characterized by \emph{non-conventional} quark and/or gluon compositions comprising \emph{multiquark} states (tetraquarks, pentaquarks, hexaquarks, heptaquarks, etc.), ``hybrid'' quark--gluon bound states, or pure-gluon bound states (nick)named glueballs.\end{itemize}There is a (crucial) \emph{fundamental} difference between conventional hadrons and exotic~hadrons, based on a (more or less) trivial observation: any colour-singlet multiquark arrangement~of~a number of quarks and/or antiquarks may be decomposed (in one or more ways)~into~a~set~of states that are also colour singlets but consist of lesser numbers of quarks and/or antiquarks.

Therefore, an (initially) tightly bound, ``compact'' \emph{multiquark} hadron may reconfigure to molecular-type clusters of (ultimately) \emph{conventional} hadrons, loosely bound by some residual forces \cite{LMS-C,HS}. In view of this, trustworthy attempts to describe exotic hadrons~should (struggle to) take into account, too, the potential mixing of these two ``phases'' of multiquark~hadrons.

The present note \emph{recalls} a collection of recently proposed procedures and considerations the application of which \emph{might facilitate} gaining theoretical understanding of (experimentally established) multiquark states. Both origin and prospects of these tools are illustrated for the hopefully easiest case: the kind of tetraquarks presumably least plagued by complications of technical nature given by (compact) bound states of two quarks and two antiquarks~carrying four unequal flavour quantum numbers. (These tools' transfer to other cases seems evident.) In particular, a brief glance at the related present \emph{experimental} situation \cite{PDG,LHCb0,LHCb1,LHCb2,LHCb3,LHCb4,LHCb5,LHCb6} (Section~\ref{sT}) will be followed by a recapitulation of insights gained upon basing the strong interactions' gauge symmetry tentatively on special unitary groups of \emph{higher} dimension \cite{GH,EW,KP,CL,MPR,LMS-N1,LMS-N2,LMS-N3,LMS-N3+,LMS-N4,LMS-N4+,LMS-N5,LMS-N5+,LMS-N6,LMS-N7} (Section~\ref{sN})~and a sketch of the advantages of trimming a popular technique for the nonperturbative analytical discussion of QCD bound states to fit to the needs of multiquark hadrons \cite{SVZ,RRY,CK,KGW,ECT1,ECT2,ECT3,ECT4,LMS-A1,LMS-A2,LMS-A3,LMS-A4,LMS-A5}~(Section~\ref{sR}).

\section{Tetraquark Mesons --- the Example of Multiquark Exotic Hadron States Par Excellence}\label{sT}All tetraquark mesons $T$ are bound states of two antiquarks $\overline q_a,\overline q_c$ and two quarks~$q_b,q_d$,\begin{equation}T=[\overline q_a\,q_b\,\overline q_c\,q_d]\ ,\qquad a,b,c,d\in\{u,d,s,c,b\}\ ,\label{tm}\end{equation}henceforth calling the masses of the four (anti-) quarks constituting such state $m_a,m_b,m_c,m_d$. On group-theoretical grounds, the presence of these mesons in the hadron spectrum without coming into conflict with \emph{confinement} of colour is rendered possible by the appearance of two SU(3) singlet representations $\mathbf{1}$ in the (appropriate) tensor product of two fundamental~SU(3) representations $\mathbf{3}$ and two (complex-conjugate) fundamental SU(3) representations $\mathbf{\overline{3}}$ \cite{LMS-N5+,LMS-A2}, as this product's decomposition into irreducible SU(3) representations $\mathbf{1},\mathbf{8},\mathbf{10},\mathbf{\overline{10}},\mathbf{27}$~reveals:\begin{equation}q_b\,q_d\,\overline q_a\,\overline q_c\sim\mathbf{3\otimes3\otimes\overline{3}\otimes\overline{3}=81=\textcolor{red}{1}\oplus\textcolor{red}{1}\oplus8\oplus8\oplus8\oplus8\oplus10\oplus\overline{10}\oplus27}\ .\label{GT}\end{equation}

As far as its \emph{flavour} degrees of freedom are concerned, the four quark constituents of any tetraquark state (\ref{tm}) may contribute, at most, four different quark flavours and, trivially,~carry at least one, the \emph{same} for all the four (anti-) quarks. Owing to such simultaneous involvement of both quarks and antiquarks, however, the latters' hadron bound states need not feature all of the available quark flavours. Table~\ref{TC} presents the listing \cite{LMS-N4} of conceivable quark-flavour arrangements in the tetraquark state (\ref{tm}), with respect to both the number of \emph{different}~flavours $a\ne b\ne c\ne d$ provided by two quarks and two antiquarks as well as the number of flavours exhibited by the related \emph{hadron}, which might differ from the former number either~because of mutual flavour--antiflavour compensations or because of quark-flavour~double occurrences.

\begin{table}[H]\caption{Tetraquark states (\ref{tm}): Classification by different vs.~open quark-flavour content~$a\ne b\ne c\ne d$, \emph{open-flavour number} referring to all flavours not counterbalanced by their antiflavours. (From Ref.~\cite{LMS-N4}.)\label{TC}}\begin{center}\begin{tabular}{ccc}\toprule\textbf{Number of Different}&$\quad$ \ \textbf{Quark Composition} \ $\quad$&\textbf{Number of Open}\\\textbf{Quark Flavours Involved}&$\overline q_\square\,q_\square\ \,\overline q_\square\,q_\square$&\textbf{Quark Flavours Involved}\\\midrule
4&$\overline q_a\,q_b\ \,\overline q_c\,q_d$&4\\\midrule
3&$\overline q_a\,q_b\ \,\overline q_c\,q_b$&4\\&$\overline q_a\,q_b\ \,\overline q_a\,q_c$&4\\&$\overline q_a\,q_b\ \,\overline q_b\,q_c$&2\\&$\overline q_a\,q_b\ \,\overline q_c\,q_c$&2\\\midrule
2&$\overline q_a\,q_b\ \,\overline q_a\,q_b$&4\\&$\overline q_a\,q_a\ \,\overline q_a\,q_b$&2\\&$\overline q_a\,q_a\ \,\overline q_b\,q_a$&2\\&$\overline q_a\,q_b\ \,\overline q_b\,q_a$&0\\&$\overline q_a\,q_a\ \,\overline q_b\,q_b$&0\\\midrule
1&$\overline q_a\,q_a\ \,\overline q_a\,q_a$&0\\\bottomrule\end{tabular}\end{center}\end{table}

Needless to say, at least from the experimental point of view it may be more satisfactory if the exotic nature of a (suspected) multiquark is established already by its observed~content of quark flavours. The corresponding species of multiquarks may be told apart by relying~on\begin{Definition}\label{dFE}A multiquark hadron is termed \textbf{flavour-exotic} if it exhibits more open quark flavours than the corresponding category of conventional hadrons does, which means at least three open~quark flavours in the case of mesonic states or at least four open quark flavours in the case of baryonic~states. By contrast, a multiquark hadron is called \textbf{flavour-cryptoexotic} if it does not meet this requirement.\end{Definition}\noindent For the quark-flavour arrangements of tetraquarks, Table~\ref{TC} offers several options to meet~the requirement of being considered flavour-exotic: of course, there can exist merely one~flavour arrangement that incorporates four mutually different quark flavours; however, there~exist a few self-evident options for \emph{flavour-exotic} tetraquarks to comprise not more than two~or three different quark flavours by involving one or even two \emph{double} appearances of a given~flavour.

Quite recently, various candidates for tetraquark states that are manifestly \emph{flavour-exotic} by exhibiting (in accordance with Definition~\ref{dFE}) four open quark flavours have been observed by experiment. Regarding the flavour compositions of these candidates, there are states each encompassing exactly one of all four lightest quarks \cite{LHCb1,LHCb2,LHCb5,LHCb6} and ``doubly~flavoured''~ones containing only three different flavours but one of these twice \cite{LHCb3,LHCb4} (see summary of Table~\ref{FlEx}).

\begin{table}[H]\caption{Flavour-exotic tetraquark states: Experimental candidates, in naming convention of LHCb~\cite{LHCb0}.\label{FlEx}}\begin{center}\begin{tabular}{lcr}\toprule
\textbf{Candidate Tetraquark Meson}&\textbf{(Minimal) Quark-Flavour Content}&\textbf{References}\\\midrule
$T_{cs0}(2900)^0$&$c\overline ds\overline u$&\cite{LHCb1,LHCb2}\\
$T_{cs1}(2900)^0$&$c\overline ds\overline u$&\cite{LHCb1,LHCb2}\\
$T_{cc}(3875)^+$&$cc\overline u\overline d$&\cite{LHCb3,LHCb4}\\
$T^a_{c\overline s0}(2900)^0$&$c\overline sd\overline u$&\cite{LHCb5,LHCb6}\\
$T^a_{c\overline s0}(2900)^{++}$&$c\overline su\overline d$&\cite{LHCb5,LHCb6}\\\bottomrule\end{tabular}\end{center}\end{table}

\section{Correlation Functions of Hadron Interpolating Operators: Application to Multiquarks}\label{CF}For descriptions of hadronic states in terms of QCD, a pivotal contact point between~the realm of QCD and the realm of hadrons is established by the concept of hadron interpolating operators. For a fixed hadron $H$ under consideration, its --- not necessarily unique --- hadron interpolating operator, generically called ${\cal O}$, is a gauge-invariant local operator~composed~of the QCD dynamical degrees of freedom, the quark and gluon field operators, that betrays its nonzero overlap with the hadron $|H\rangle$ by the nonvanishing matrix element emerging from its getting sandwiched between the hadronic state $|H\rangle$ and the QCD vacuum $|0\rangle$: $\langle0|{\cal O}|H\rangle\ne0$. In all subsequent implementations of hadron interpolating operators, features such as parity or spin degrees of freedom can be safely ignored; they get therefore notationally suppressed.

For a \emph{conventional} meson consisting of a quark of flavour $b$ and an antiquark of~flavour $a$, the most evident option for its interpolating operator is the quark--antiquark bilinear~current\begin{equation}j_{\overline ab}(x)\equiv\overline q_a(x)\,q_b(x)\ .\label{b}\end{equation}For \emph{exotic} hadrons belonging to the subset of \emph{tetraquark} mesons characterized in Equation~(\ref{tm}), the search for appropriate tetraquark interpolating operators, specifically named $\theta$, is greatly facilitated by the observation \cite{RLJ} that (by means of suitable Fierz transformations~\cite{MF}) \emph{every} colour-singlet operator that is composed of two quarks and two antiquarks can~be expressed by a linear combination of only two different products of colour-singlet conventional-meson interpolating operators of quark-bilinear-current shape (\ref{b}). Thus, this ``operator basis'' reads\begin{equation}\theta_{\overline ab\overline cd}(x)\equiv j_{\overline ab}(x)\,j_{\overline cd}(x)\ ,\qquad\theta_{\overline ad\overline cb}(x)\equiv j_{\overline ad}(x)\,j_{\overline cb}(x)\ .\label{t}\end{equation}Moreover, taking into account some useful identities recalled, for instance, by Equations~(32) and (36) of Reference~\cite{LMS-N7} or Equations (1) and (2) of Reference~\cite{LMS-A3} \emph{may} be regarded either as a kind of shortcut to or as explicit verification of these findings. The tetraquark interpolating operators (\ref{t}) will provide some kind of playground for (most of) the ensuing considerations.

That pleasing observation \cite{RLJ} points out a promising route how to reasonably proceed. Namely, the enabled basic two-current structure (\ref{t}) of the tetraquark interpolating operators $\theta$ suggests to start (envisaged) analyses of tetraquarks from \emph{correlation functions} --- in~general, defined by vacuum expectation values of time-ordered products, symbolized by T, of chosen field operators --- of four quark-bilinear operators (\ref{b}). If tolerated by the involved dynamics, in appropriate four-point correlation functions of such kind tetraquark states should become manifest by their contributions in form of intermediate-state poles. Momentarily focusing to only essential aspects, all these four-current correlation functions are of the general~structure\begin{equation}\left\langle{\rm T}\!\left(j(y)\,j(y')\,j^\dag(x)\,j^\dag(x')\right)\right\rangle\ .\label{4}\end{equation}Upon application of well-understood procedures, the correlation functions (\ref{4}) entail also the amplitudes encoding scatterings of two conventional mesons into two conventional~mesons. Because of the two-current structure (\ref{t}), contact with tetraquark states, in form of correlation functions involving tetraquark interpolating operators $\theta$, can be established by identification or contraction of configuration-space coordinates of \emph{proper} quark-bilinear currents $j$,~forming\begin{itemize}\item\emph{twice} configuration-space contracted \emph{two-point} correlation functions of \emph{two} operators~(\ref{t})\begin{equation}\left\langle{\rm T}\!\left(\theta(y)\,\theta^\dag(x)\right)\right\rangle=\lim_{\underset{\scriptstyle y'\to y}{\scriptstyle x'\to x}}\left\langle{\rm T}\!\left(j(y)\,j(y')\,j^\dag(x)\,j^\dag(x')\right)\right\rangle\ ;\label{2}\end{equation}\item\emph{once} contracted \emph{three-point} correlation functions of \emph{one} operator (\ref{t}) and \emph{two} operators~(\ref{b})\begin{equation}\left\langle{\rm T}\!\left(j(y)\,j(y')\,\theta^\dag(x)\right)\right\rangle=\lim_{x'\to x}\left\langle{\rm T}\!\left(j(y)\,j(y')\,j^\dag(x)\,j^\dag(x')\right)\right\rangle\ .\label{3}\end{equation}\end{itemize}

An immediate implication of the mere conceptual nature of unconventional multiquark states is, as already stressed in Section~\ref{sE}, their potential to undergo \emph{clustering} without getting into conflict with colour confinement \cite{LMS-C}. For the correlation-function underpinned analyses of tetraquark properties, this finding should be regarded as a strong hint that, presumably or even very likely, not all QCD-level contributions to some correlation function are, in general, of relevance for such formation of a tetraquark pole. It appears opportune to distinguish~any contribution that may play a r\^ole in tetraquark studies even by nomenclature; this is done~in\begin{Definition}A QCD contribution to a correlation function (\ref{4}) is termed \textbf{tetraquark-phile}~\cite{LMS-N2,LMS-N5} if it is (potentially) capable of supporting the formation of a tetraquark-related intermediate-state~pole.\label{dTP}\end{Definition}\noindent As a guidance through the process of filtering all of the QCD-level contributions as implicitly requested by Definition~\ref{dTP}, a self-evident, easy to implement criterion may be devised \cite{LMS-N1,LMS-N3}:\begin{Proposition}\label{pTP}For a given four-point correlation function (\ref{4}) with external momenta in initial~state $p_1$, $p_2$ and external momenta in final state $q_1$, $q_2$, considered as a function of the Mandelstam variable\begin{equation}s\equiv(p_1+p_2)^2=(q_1+q_2)^2\ ,\end{equation}a QCD-level contribution is supposed to be tetraquark-phile if it exhibits a nonpolynomial dependence on $s$ and if it develops an intermediate-state four-quark-related branch cut starting at the branch~point\begin{equation}\hat s\equiv(m_a+m_b+m_c+m_d)^2\ .\end{equation}\end{Proposition}\noindent For any contribution to a correlation function, the capability of supporting the formation of a tetraquark pole by satisfying all requirements in Proposition~\ref{pTP} may be straightforwardly and unambiguously decided by consulting the related Landau equations \cite{LDL}: the existence~of an appropriate solution to (the relevant set of) those Landau equations indicates the presence~of an expected branch cut. References~\cite{LMS-N3,LMS-N7,LMS-A3} show some examples worked out in all details.

\begin{figure}[H]\includegraphics[width=10.633cm]{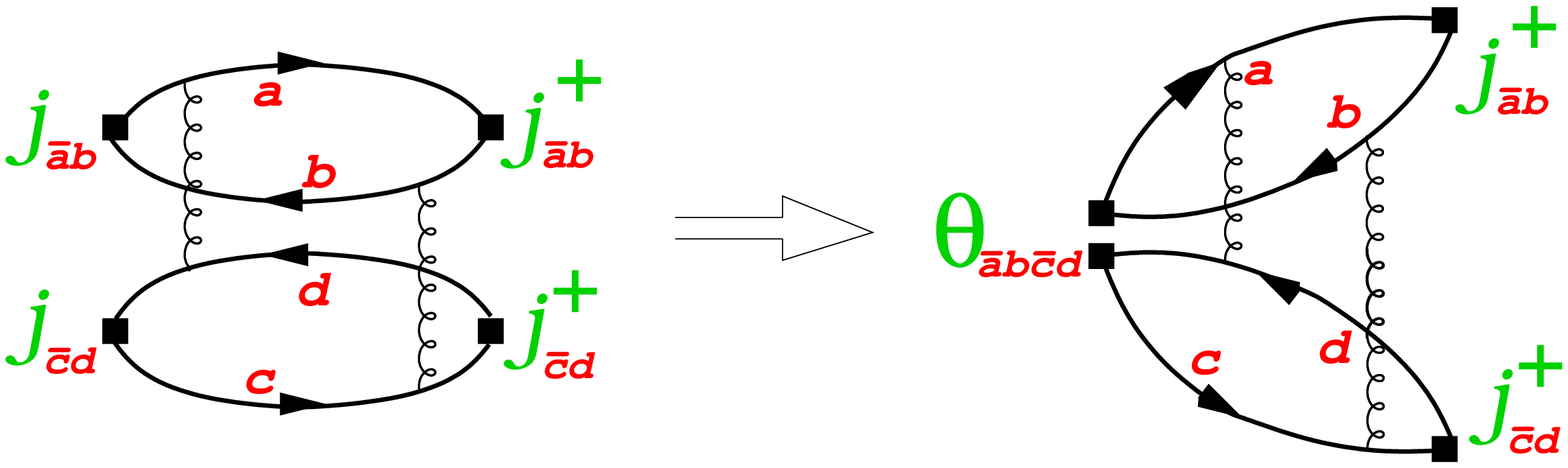}\caption{Definitely flavour-exotic four-current correlation function (\ref{4}) of flavour-preserving type~(left) and (right) contraction (\ref{3}) to a correlation function of one tetraquark interpolating operator~(\ref{t}) and two quark-bilinear currents (\ref{b}) \cite{LMS-A1}: typical contribution of \emph{lowest tetraquark-phile} perturbative order~$O(\alpha_{\rm s}^2)$.\label{F3p}}\end{figure}

\begin{figure}[H]\includegraphics[width=10.633cm]{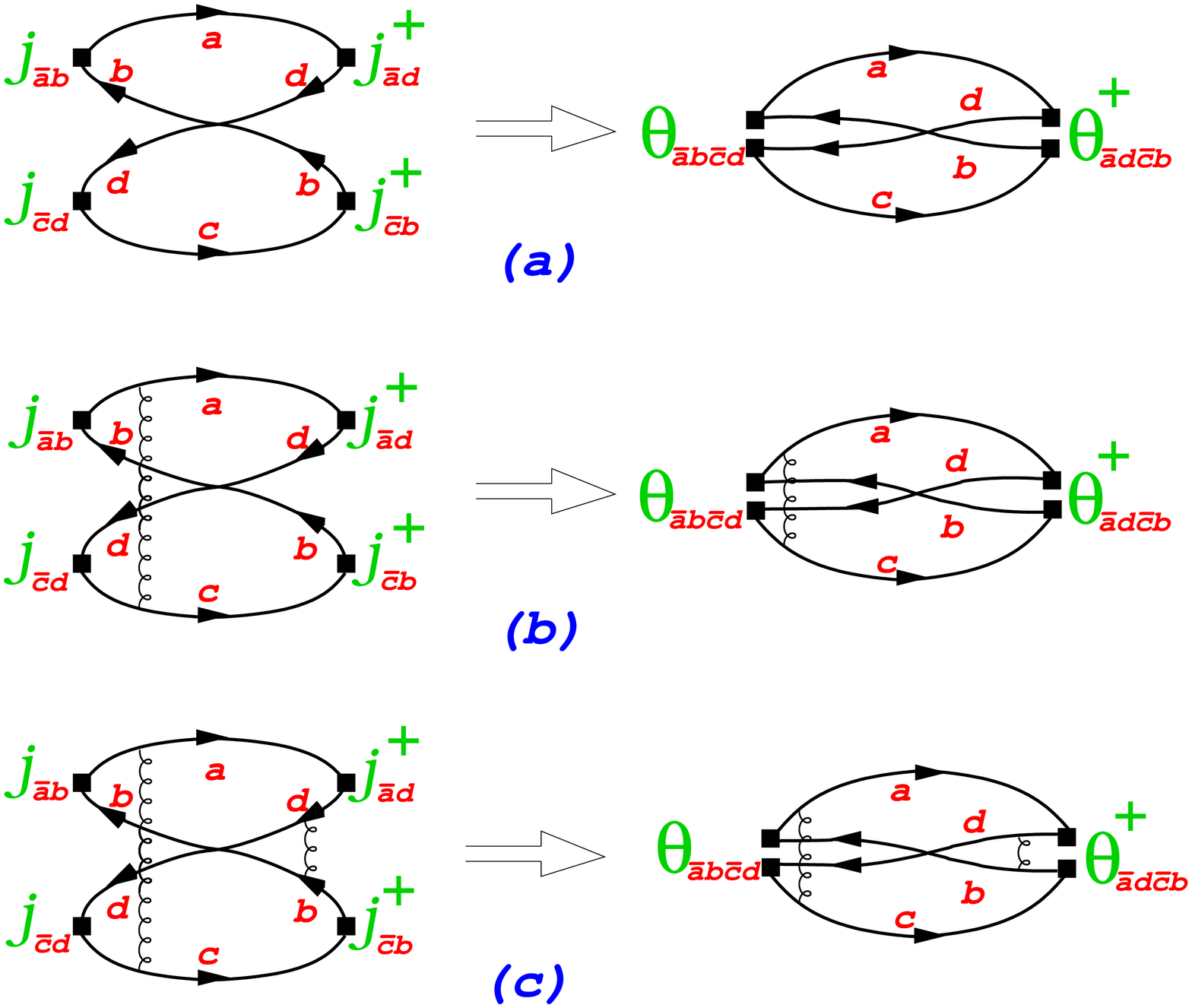}\caption{Definitely flavour-exotic four-current correlation function (\ref{4}) of flavour-\emph{rearranging} type~(left) and (right) its contraction (\ref{2}) to two-point correlation function of tetraquark interpolating operators~(\ref{t}) \cite{LMS-A1,LMS-A3}. Representative contributions of lowest perturbative orders: (\textbf{a}) $O(\alpha_{\rm s}^0)$, (\textbf{b}) $O(\alpha_{\rm s})$ and (\textbf{c})~$O(\alpha_{\rm s}^2)$.\label{F2r}}\end{figure}

\begin{figure}[H]\includegraphics[width=10.633cm]{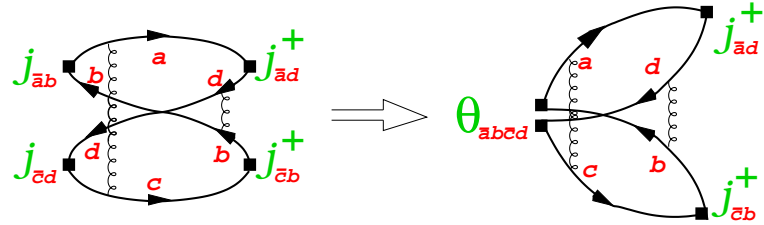}\caption{Definitely flavour-exotic four-current correlation function (\ref{4}) of flavour-\emph{rearranging} type~(left) and (right) contraction (\ref{3}) to a correlation function of one tetraquark interpolating operator~(\ref{t}) and two quark-bilinear currents (\ref{b}) \cite{LMS-A3}: typical contribution of \emph{lowest tetraquark-phile} perturbative order~$O(\alpha_{\rm s}^2)$.\label{F3r}}\end{figure}

As announced in Section~\ref{sE}, the benefit of implementing such programme is exemplified for the meanwhile even experimentally observed \cite{LHCb1,LHCb2,LHCb5,LHCb6} \emph{subset} of all those~flavour-exotic tetraquarks that exhibit not less than (the feasible maximum of) four unequal~quark~flavours:\begin{Definition}The quark-flavour composition of a tetraquark (\ref{tm}) is called \mbox{\textbf{definitely flavour-exotic}} if it comprises four mutually different quark flavours $a\ne b\ne c\ne d$, that is, if this state is of~the~kind\begin{equation}T=[\overline q_a\,q_b\,\overline q_c\,q_d]\ ,\qquad a,b,c,d\in\{u,d,s,c,b\}\ ,\qquad a\ne b\ne c\ne d\ .\label{F}\end{equation}\end{Definition}\noindent At least for the case of the definitely flavour-exotic tetraquarks (\ref{F}), there exist two definitely distinguishable quark-flavour distributions in (from the point of view of intermediate states) incoming and outgoing states of a correlation function (\ref{4}): its quark-flavour arrangements~in initial and final state might be either identical or different. These two possibilities~got names:\begin{Definition}\label{dPR}A definitely flavour-exotic correlation function (\ref{4}) of four interpolating currents~(\ref{b})~is\begin{itemize}\item\textbf{flavour-preserving} \cite{LMS-N3+} for equal quark-flavour distributions of incoming and outgoing~states,\item\textbf{flavour-rearranging} \cite{LMS-N3+} for unlike incoming- and outgoing-state quark-flavour distributions.\end{itemize}\end{Definition}\noindent For the two categories of correlation functions (\ref{4}), it is straightforward yet worthwhile (since instructive) to investigate their contributions of lowest orders to the perturbative expansions in powers of the strong fine-structure constant (\ref{as}). Representative examples of contributions are given, for flavour-preserving cases, in Figures~\ref{F2p} and \ref{F3p} and, for flavour-rearranging cases, in Figures~\ref{F2r} and \ref{F3r}. (In the plots, internal gluon exchanges are depicted~in~form of curly lines.) Expectably, such considerations disclose differences in analyses but similarities~in~outcomes:
\begin{figure}[H]\includegraphics[width=10.633cm]{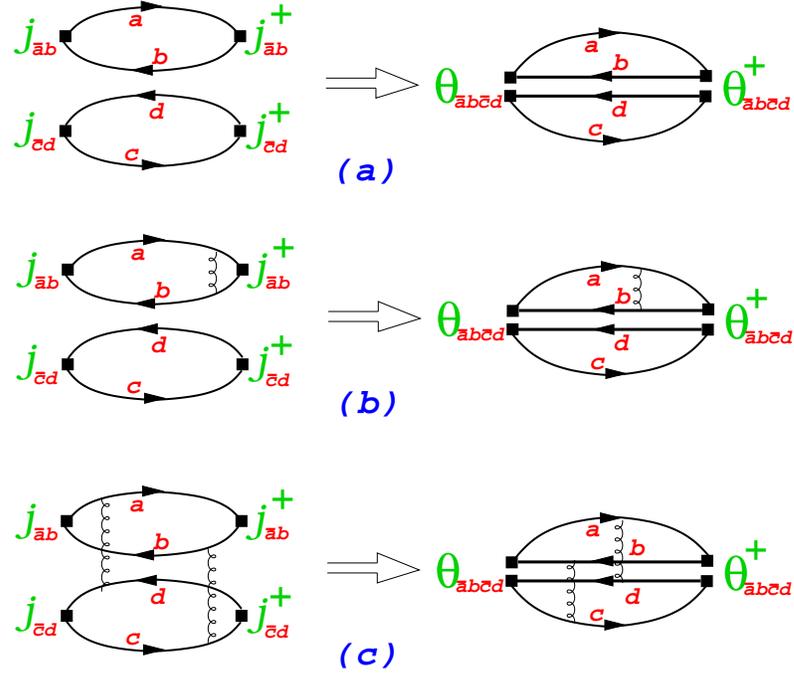}\caption{Definitely flavour-exotic four-current correlation function (\ref{4}) of flavour-preserving type~(left) and (right) its contraction (\ref{2}) to two-point correlation function of tetraquark interpolating operators~(\ref{t}) \cite{LMS-A1,LMS-A4}. Representative contributions of lowest perturbative orders: (\textbf{a}) $O(\alpha_{\rm s}^0)$, (\textbf{b}) $O(\alpha_{\rm s})$ and (\textbf{c})~$O(\alpha_{\rm s}^2)$.\label{F2p}}\end{figure}
\begin{itemize}\item For flavour-preserving correlation functions, the line of argument proves to be, more~or less, evident. \emph{All} the contributions of the type of Figure~\ref{F2p}(a) or of the type of Figure~\ref{F2p}(b), involving at most one gluon exchange, are doubtlessly disconnected. The contributions that involve a single gluon exchange between their two (otherwise disconnected)~quark loops \emph{vanish} identically, due to the vanishing of the sum over colour degrees of freedom of each of the two quark loops. Phrased \emph{slightly} more technically, this can be~traced~back to the tracelessness of all generators of a special unitary group, governing the couplings of quarks and gluons. Consequently, \emph{exclusively} contributions that involve, at least, two gluon exchanges of an appropriate topology may be viewed as tetraquark-phile. These insights get, of course, corroborated by identifying these tetraquark-phile contributions according to Proposition~\ref{pTP} by explicit inspection \cite{LMS-N1} by way of their Landau equations. Replacing any double contraction~(\ref{2}) in Figure~\ref{F2p} by a single contraction~(\ref{3}) confirms the tetraquark-phile nature of contributions of the type of Figure~\ref{F3p} or related higher~orders.\item For flavour-rearranging correlation functions, a simple optical guidance in this analysis is, beyond doubt, hardly imaginable: already the lowest-order contributions~turn~out~to be connected. Rather, one has to gladly accept any assistance offered by that tool~called Landau equations. For the three lowest-order contributions exemplified in Figure~\ref{F2r},~the usage of this formalism is demonstrated, in full detail, in Appendix~A of Reference~\cite{LMS-N3}, in the Appendix of Reference~\cite{LMS-A3}, as well as in Section~4 of Reference~\cite{LMS-N7}. For this~kind of analysis, it might prove advantageous to recast the encountered plots into box shape, by ``unfolding'' all these plots \cite{CL,LMS-N3,LMS-A3,LMS-N7}. These efforts' outcome is that contributions of the type of Figure~\ref{F2r}(a) or of the type of Figure~\ref{F2r}(b), being characterized by no or~only one internal gluon exchange, do not incorporate the requested four-quark~singularities. Involvement of this feature starts not before the level of two gluon exchanges~of~\emph{suitable} positioning, which then holds, of course, also for the single~contractions~(\ref{3}) in Figure~\ref{F3r}.\end{itemize}As an overall summary of the two classes of definitely flavour-exotic correlation functions~(\ref{4}) identified by Definition~\ref{dPR}, the systematic scrutiny of their lowest-order contributions~betrays that tetraquark-phile contributions (an essential ingredient, since providing the singularities that, upon summation, \emph{may} support the development of intermediate-state tetraquark~poles) will not emerge before the next-to-next-to-lowest order in a series expansion in powers~of~the strong fine-structure constant (\ref{as}), that is, in terms of $\alpha_{\rm s}$, have to be at least of the order~$O(\alpha_{\rm s}^2)$.

\section{Number of Colour Degrees of Freedom, Unfixed: Large-$N_{\rm c}$ Limit and $1/N_{\rm c}$ Expansion}\label{sN}Quite generally, first insights, even if only of qualitative nature, may be gained from the reduction of the complexity of QCD, enacted by the increase of the number of colour~degrees of freedom and, in parallel, the decrease of the strength of the strong-interaction coupling~$g_{\rm s}$. In some more detail, that simplification of QCD \cite{GH,EW} proceeds along the following moves:\begin{itemize}\item Generalize QCD to the gauge theories invariant under a non-Abelian Lie group~SU($N_{\rm c}$). The dynamical degrees of freedom of each of the latter quantum field theories~hence~are its gauge bosons, still retaining their designation as gluons and transforming according to the $(N_{\rm c}^2-1)$-dimensional, adjoint representation of SU($N_{\rm c}$), and its~fermionic~quarks that transform according to the $N_{\rm c}$-dimensional, fundamental representation of SU($N_{\rm c}$).\item Allow the number of colour degrees of freedom, $N_{\rm c}$, to increase from $N_{\rm c}=3$~to~infinity:\begin{equation}N_{\rm c}\to\infty\ .\label{lnl}\end{equation}\item For the strong coupling strength $g_{\rm s}$, demand the \emph{related} decrease, with rising $N_{\rm c}$,~to~zero:\begin{equation}g_{\rm s}\propto\frac{1}{\sqrt{N_{\rm c}}}=O(N_{\rm c}^{-1/2})\xrightarrow[N_{\rm c}\to\infty]{}0\ .\label{LNLg}\end{equation}Clearly, for the strong fine-structure coupling $\alpha_{\rm s}$ this requirement implies the behaviour\begin{equation}\alpha_{\rm s}\propto\frac{1}{N_{\rm c}}=O(N_{\rm c}^{-1})\xrightarrow[N_{\rm c}\to\infty]{}0\ .\label{LNLa}\end{equation}Therefore, in the large-$N_{\rm c}$ limit, the product $N_{\rm c}\,\alpha_{\rm s}$ approaches a meaningful finite~value.\end{itemize}Only by establishing a careful balance between the growth of $N_{\rm c}$ and the vanishing of~$\alpha_{\rm s}$, the latter requirement allows for both reasonable generalization of QCD to its large-$N_{\rm c}$~limit and exploitation of any corresponding $1/N_{\rm c}$ expansion, that is, the expansion in powers~of~$1/N_{\rm c}$.

According to the above characterization of large-$N_{\rm c}$ QCD, for each QCD contribution to a correlation function its behaviour in the large-$N_{\rm c}$ limit gets determined by two ingredients:\begin{itemize}\item the number of \emph{closed} loops of the colour degrees of freedom carried by quarks or~gluons,\item the number of either the strong couplings (\ref{LNLg}) or the strong fine-structure constants~(\ref{LNLa}).\end{itemize}Keeping this in mind, the large-$N_{\rm c}$ behaviour of arbitrary correlation functions will be found. In particular, for the tetraquark-phile (and therefore tetraquark-pole relevant) contributions, indicated by the subscript ``tp'', to definitely flavour-exotic correlation functions~(\ref{4}), one~gets\begin{itemize}\item for any flavour-preserving contribution of the type employed by Figure~\ref{F2p}(c) or~Figure~\ref{F3p},\begin{align}&\left\langle{\rm T}\!\left(j_{\overline ab}(y)\,j_{\overline cd}(y')\,j^\dag_{\overline ab}(x)\,j^\dag_{\overline cd}(x')\right)\right\rangle_{\rm tp}=O(N_{\rm c}^2\,\alpha_{\rm s}^2)=O(N_{\rm c}^0)\ ,\label{fpt1}\\[.23ex]&\left\langle{\rm T}\!\left(j_{\overline ad}(y)\,j_{\overline cb}(y')\,j^\dag_{\overline ad}(x)\,j^\dag_{\overline cb}(x')\right)\right\rangle_{\rm tp}=O(N_{\rm c}^2\,\alpha_{\rm s}^2)=O(N_{\rm c}^0)\ ,\label{fpt2}\end{align}\item for each flavour-rearranging contribution of the kind adopted by Figure~\ref{F2r}(c) or Figure~\ref{F3r},
\begin{equation}\left\langle{\rm T}\!\left(j_{\overline ab}(y)\,j_{\overline cd}(y')\,j^\dag_{\overline ad}(x)\,j^\dag_{\overline cb}(x')\right)\right\rangle_{\rm tp}=O(N_{\rm c}\,\alpha_{\rm s}^2)=O(N_{\rm c}^{-1})\ .\label{frt}\end{equation}\end{itemize}

This general discrepancy between the large-$N_{\rm c}$ behaviour of the flavour-preserving and of the flavour-rearranging four-point correlation functions expressed, for all contributions of any tetraquark-phile type, by Equations~(\ref{fpt1}) and (\ref{fpt2}), on the one hand, and by Equation~(\ref{frt}), on the other hand, has a startling or even disturbing implication for the spectra of tetraquark mesons to be expected in the large-$N_{\rm c}$ limit. In the scattering of a pair of \emph{conventional} mesons,\begin{equation}M_{\overline ab}=[\overline q_a\,q_b]\ ,\qquad a,b\in\{u,d,s,c,b,t(,\dots?)\}\ ,\label{cm}\end{equation}a tetraquark $T$ betrays its existence by contributing in form of an intermediate-state pole. Its couplings to conventional mesons are governed by \emph{transition amplitudes} $A(T\longleftrightarrow M_{\overline ab}\,M_{\overline cd})$. Given the discrepancy between those classes of contributions for large $N_{\rm c}$, consistency in the large-$N_{\rm c}$ limit turns out \cite{LMS-N1,LMS-N3} to impose constraints on any involved transition~amplitudes.

The QCD predictions for the large-$N_{\rm c}$ behaviour of the correlation functions introduced in Section~\ref{CF} cannot be matched, at hadron level, by the presence of merely a single tetraquark state \cite{LMS-N4+}. Rather, fulfillment of the large-$N_{\rm c}$ behaviour requested by Equations~(\ref{fpt1}), (\ref{fpt2}) and (\ref{frt}) by the tetraquark-pole contributions necessitates the pairwise occurrence of tetraquarks, that is to say, of a minimum of two (corresponding) tetraquarks \cite{LMS-N1,LMS-N3}. The two tetraquarks, generically denoted by $T_A$ and $T_B$, have to exhibit \emph{unequal} $N_{\rm c}$ dependences of their transition amplitudes to the two possible quark-flavour divisions among the two conventional mesons in initial and final states; their dominant decay channels, however, exhibit the same large-$N_{\rm c}$ behaviour. Thus, in the large-$N_{\rm c}$ limit their total decay widths, $\Gamma$, behave in a similar~fashion,\begin{equation}\Gamma(T_A)=O(N_{\rm c}^{-2})=\Gamma(T_B)\ ,\end{equation}and the large-$N_{\rm c}$ interrelationships of the four involved transition amplitudes are of the~kind
\begin{align}\underbrace{A(T_A\longleftrightarrow M_{\overline ab}\,M_{\overline cd})=O(N_{\rm c}^{-1})}
_{\mbox{$\Longrightarrow\qquad\Gamma(T_A)=O(N_{\rm c}^{-2})$}}\quad\stackrel{\fbox{$N_{\rm c}$ order}}{>}\quad
A(T_A\longleftrightarrow M_{\overline ad}\,M_{\overline cb})=O(N_{\rm c}^{-2})\ ,&\\
A(T_B\longleftrightarrow M_{\overline ab}\,M_{\overline cd})=O(N_{\rm c}^{-2})\quad\stackrel{\fbox{$N_{\rm c}$ order}}{<}\quad
\underbrace{A(T_B\longleftrightarrow M_{\overline ad}\,M_{\overline cb})=O(N_{\rm c}^{-1})}
_{\mbox{$\Longrightarrow\qquad\Gamma(T_B)=O(N_{\rm c}^{-2})$}}\ .&\end{align}Table~\ref{W} compares several available \emph{expectations} for the large-$N_{\rm c}$ dependence of the total decay rates $\Gamma$ of definitely exotic and cryptoexotic tetraquarks, indicating a few discrepancies~likely resulting from differences in underlying assumptions or contributions considered as crucial.

\begin{table}[H]\caption{Tetraquark total decay widths: expected upper bounds on large-$N_{\rm c}$ behaviour (from~Ref.~\cite{LMS-N4}).\label{W}}\begin{center}\begin{tabular}{lccr}\toprule\textbf{Author Collective}&\multicolumn{2}{c}{\textbf{Decay Width $\Gamma$}}&\textbf{References}\\&$\quad$\textbf{Definitely Exotic}$\quad$&$\quad$\textbf{Cryptoexotic}$\quad$&\\
&\textbf{Tetraquarks}&\textbf{Tetraquarks}&\\\midrule
Knecht, Peris&$O(1/N_{\rm c}^2)$&$O(1/N_{\rm c})$&\cite{KP}\\Cohen, Lebed&$O(1/N_{\rm c}^2)$&---&\cite{CL}\\
Maiani, Polosa, Riquer&$O(1/N_{\rm c}^3)$&$O(1/N_{\rm c}^3)$&\cite{MPR}\\
Lucha, Melikhov, Sazdjian&$O(1/N_{\rm c}^2)$&$O(1/N_{\rm c}^2)$&\cite{LMS-N1,LMS-N3}\\\bottomrule\end{tabular}\end{center}\end{table}

\section{Multiquark-Adequate QCD Sum Rules Recognizing ``Peculiarities'' of Exotic Hadrons}\label{sR}From a mainly theoretical point of view, the description of any hadronic bound~states of the fundamental degrees of freedom of QCD in a thoroughly analytical fashion appears to be most favourable; a promising approach complying with this intention, well-grounded in the framework of relativistic quantum field theories, is realized by the QCD sum rule~formalism.

In the version originally devised by Shifman, Vainshtein, Zakharov \cite{SVZ} and others \cite{RRY}, a QCD sum rule embodies an analytical relationship between, on the one hand,~properties of the hadron state (formed by the strong interactions) in the focus of one's current interest and, on the other hand, the (few) \emph{basic} parameters of their underlying quantum field theory,~QCD. In principle, every \emph{routine} derivation of a QCD sum rule follows meanwhile well-established procedures \cite{CK}. The starting point of the construction of a QCD sum rule is the evaluation of an appropriate correlation function --- which clearly has to involve an operator interpolating the hadron under investigation --- in parallel both at the phenomenological hadron level and at the fundamental QCD level, followed (of course) by equating both evaluations'~outcomes:\begin{itemize}\item In the course of QCD-level evaluation, Wilson's \emph{operator product expansion}~\cite{KGW}~(enabling conversion of a nonlocal product of operators into a series of local operators) is invoked to separate nonperturbative and (to some extent calculable) perturbative contributions.\begin{itemize}\item The \emph{perturbative} contributions, identical to the lowest term of this operator product expansion, can be inferred in form of a series in powers of the strong coupling~(\ref{as}).\item The \emph{nonperturbative} contributions involve, apart from derivable prefactors, \emph{vacuum condensates}, i.e., the vacuum expectation values of products of quark and/or~gluon field operators, which may be interpreted as a kind of effective parameters~of~QCD.\end{itemize}\item In the course of hadron-level evaluation, the insertion of a complete set of hadron~states guarantees that the hadron under study shows up by way of its intermediate-state~pole.\end{itemize}By application of dispersion relations (and, if necessary, a sufficient number of \mbox{subtractions}), both perturbative QCD-level evaluation and hadron-level evaluation can be reexpressed~(for the sake of convenience) in the form of dispersion integrals of appropriate spectral densities.

The predictive value and therefore usefulness of the QCD--hadron relations constructed in this manner is perceptibly increased by taking consecutively both the following~measures:\begin{enumerate}\item Subject both sides of such a relation to a Borel transformation to another variable~called Borel parameter $\tau$. This results in the \emph{entire} removal of any subtraction term introduced and the suppression of the hadron-level contributions above the hadronic~ground~state. Under a Borel transformation, all \emph{vacuum condensates} in the nonperturbative QCD-level contributions get multiplied by powers of $\tau$. So, these terms are called \emph{power~corrections}.\item Rely on the assumption of quark--hadron duality, which postulates a (needless~to~stress, approximately realized) cancellation of all perturbative QCD-level contributions~above suitably defined effective thresholds, $s_{\rm eff}$, against all higher hadron-level contributions, consisting of hadron excitations and hadron continuum. In implementing~this~concept, the problem of pinning down the nature of $s_{\rm eff}$ may be dealt with in two different~ways:\begin{itemize}\item Without knowing better, just a guessed \emph{fixed} value of the parameter~$s_{\rm eff}$ is adopted:\begin{equation}s_{\rm eff}=\mbox{const}\ .\end{equation}\item In contrast, slipping in \emph{limited} information about a targeted hadron state opens the possibility \cite{ECT1,ECT2,ECT3,ECT4} to work out the \emph{expected} $s_{\rm eff}$ dependence on Borel parameters~$\tau$:\begin{equation}s_{\rm eff}=s_{\rm eff}(\tau)\ .\end{equation}\end{itemize}\end{enumerate}

The roadmap for the construction of QCD sum rules sketched above has originally been drafted for analyses of \emph{conventional} hadrons. Its unreflected application (in unchanged form) also to multiquark states seems, in view of the far-reaching discrepancies between the exotic and the conventional categories of hadrons, to be either too optimistic or a little bit too na\"ive. Rather, one should be open for (potentially favourable) modifications of the customary~QCD sum-rule approach, modifications that might be capable of improving the achieved accuracy of the predictions of QCD sum rules for the class of multiquark exotic~hadrons. In particular, upon performing necessary evaluations of correlation functions at QCD level one might find advantageous to take into account the QCD contributions' feature of being tetraquark-phile, in Definition~\ref{dTP} implied to be desirable and by Proposition~\ref{pTP} given its precise meaning, or not. With respect to the power corrections, in any QCD sum-rule derivation indispensable for its QCD-level evaluation, the problem of whether a given nonperturbative vacuum-condensate contribution is tetraquark-phile or not may be analyzed along the lines indicated in Section~\ref{CF} (as has been demonstrated at the example of definitely flavour-exotic tetraquarks \cite{LMS-A1,LMS-A3,LMS-N6}).

Targeting \emph{definitely flavour-exotic tetraquarks} (\ref{F}), the versions of correlation functions (\ref{4}) indicated in Definition \ref{dPR} have to be discriminated and hence subjected to separate~treatment.\begin{itemize}\item In the flavour-preserving case, one has to start from the four-point correlation functions\begin{equation}\left\langle{\rm T}\!\left(j_{\overline ab}(y)\,j_{\overline cd}(y')\,j^\dag_{\overline ab}(x)\,j^\dag_{\overline cd}(x')\right)\right\rangle\ ,\qquad\left\langle{\rm T}\!\left(j_{\overline ad}(y)\,j_{\overline cb}(y')\,j^\dag_{\overline ad}(x)\,j^\dag_{\overline cb}(x')\right)\right\rangle\ .\label{fp}\end{equation}Applying the \emph{traditional} QCD sum-rule manipulations to twofold contractions (\ref{2}) of the correlation functions (\ref{fp}) yields as outcome of this enterprise a relationship,~depicted in Figure~\ref{FQSR+p}, that incorporates a (vast) multitude of QCD-level and hadron-level quantities.\begin{figure}[H]\includegraphics[width=13.87cm]{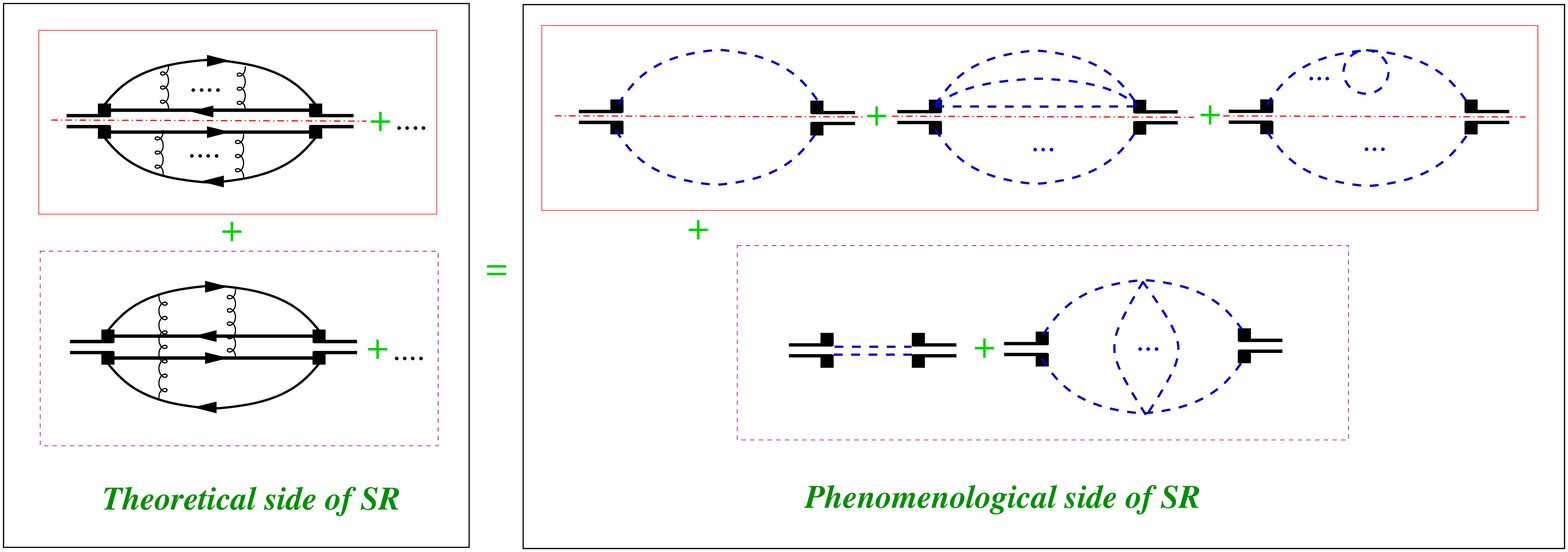}\caption{Aggregation of a pair of \emph{unconnected} conventional-meson QCD sum rules of the kind recalled by Figure~\ref{FQSRC} (top row, separated by a red dot-dashed line) and (bottom row) the \emph{tetraquark-adequate} QCD sum rule of generic structure as in Figure~\ref{FQSRT}, potentially supporting tetraquark intermediate-state poles: outcome of uncritical evaluation of correlation functions (\ref{fp}) still awaiting its disentanglement~\cite{LMS-A1,LMS-A4}.\label{FQSR+p}}\end{figure}However, a more in-depth analysis \cite{LMS-A1} reveals that, already on diagrammatic grounds, this conglomerate decomposes, in fact, into two QCD sum rules for \emph{conventional}~mesons (Figure~\ref{FQSRC}) and one further QCD sum rule that, potentially, supports the development of \emph{tetraquark poles} and rightly deserves the label of being ``tetraquark-adequate'' (Figure~\ref{FQSRT}). In the course of its QCD-level evaluation, this latter QCD sum rule receives, exclusively, tetraquark-phile contributions, in the sense of Proposition~\ref{pTP}; all the perturbative among these enter in form of dispersion integrals of tetraquark-adequate spectral~densities,~$\rho_{\rm p}$. An analogous reflection for single contractions (\ref{3}) of the correlation functions (\ref{fp}) leads to similar QCD sum-rule findings, all perturbative tetraquark-phile QCD contributions being encoded, in dispersive formulation, in tetraquark-adequate spectral densities~$\Delta_{\rm p}$.\begin{figure}[H]\includegraphics[width=10.254cm]{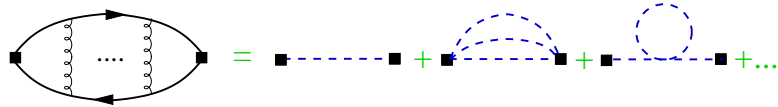}\caption{Schematical composition of QCD sum rules for conventional mesons (blue dashed lines)~\cite{LMS-A1}.\label{FQSRC}}\end{figure}\begin{figure}[H]\includegraphics[width=10.254cm]{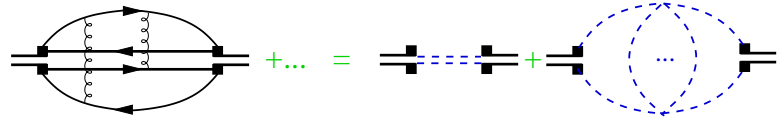}\caption{Schematical composition of a tetraquark-adequate QCD sum rule of flavour-preserving type: tetraquark-phile contributions at QCD level, at hadron level counterbalanced by non-separable~meson contributions (blue dashed lines), and perhaps those of tetraquark poles (blue dashed double line)~\cite{LMS-A1}.\label{FQSRT}}\end{figure}\item In the flavour-rearranging case, one has to deal with the four-point correlation function\begin{equation}\left\langle{\rm T}\!\left(j_{\overline ab}(y)\,j_{\overline cd}(y')\,j^\dag_{\overline ad}(x)\,j^\dag_{\overline cb}(x')\right)\right\rangle\ .\label{fr}\end{equation}Here, irrespective of (ultimately necessary) spatial contractions (\ref{2}) and (\ref{3}) of four-point correlation functions (\ref{4}), the analysis is unfortunately not thus straightforward as in the flavour-preserving case: Within QCD-level evaluation, all \emph{tetraquark-phile} contributions (defined by requiring them to satisfy the constraint formulated in Proposition~\ref{pTP})~may~be identified, case by case, by inspection of the solutions of the relevant Landau equations. Within hadron-level evaluation, that QCD-level characteristic of being tetraquark-phile or not is mirrored by the ability of any contributions at hadron level to accommodate, in their $s$ channel, two-meson intermediate states or not, in addition to a possible presence of tetraquark intermediate-state poles \cite{LMS-A3}. Hardly surprisingly, these insights translate the outcome of the QCD sum-rule formalism based on the correlation function (\ref{fr}) into a quark--hadron relation of (expected) two-component structure symbolically~shown in Figure~\ref{FQSR+r}. All \emph{perturbative} tetraquark-phile QCD-level contributions find their way into a tetraquark-adequate QCD sum rule arising from a \emph{precursor} as in Figure~\ref{FQSR+r}(b) by~spectral densities $\rho_{\rm r}$ in the double-contractions case (\ref{2}) and $\Delta_{\rm r}$ in the single-contraction case~(\ref{3}).\begin{figure}[H]\includegraphics[width=10.633cm]{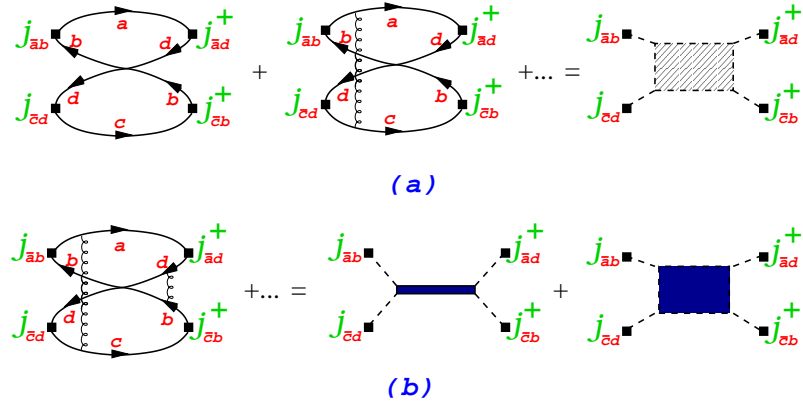}\caption{Outcome of application of established QCD sum-rule techniques to correlation functions (\ref{fr}), consisting of two uncorrelated quark--hadron relationships: (\textbf{a}) one equating the non-tetraquark-phile QCD contributions with hadron contributions not involving any two-meson $s$-channel cuts~(subsumed by hatched rectangle); (\textbf{b}) the precursor of a tetraquark-adequate QCD sum rule, involving \mbox{two-meson} $s$-channel cuts (subsumed by filled rectangle) and \emph{maybe} tetraquark poles (blue horizontal bar) too~\cite{LMS-A3}.\label{FQSR+r}}\end{figure}\end{itemize}For a definitely flavour-exotic tetraquark (\ref{F}), the properties of foremost interest are~mass~$M$,\begin{itemize}\item decay constants $f_{\overline ab\overline cd}$ and $f_{\overline ad\overline cb}$, arising from the vacuum--tetraquark matrix elements~of the two distinct operators (\ref{t}) interpolating any definitely flavour-exotic tetraquark~(\ref{F}),\begin{equation}f_{\overline ab\overline cd}\equiv\langle0|\theta_{\overline ab\overline cd}|T\rangle\ ,\qquad f_{\overline ad\overline cb}\equiv\langle0|\theta_{\overline ad\overline cb}|T\rangle\ ;\label{f}\end{equation}\item momentum-space amplitudes $A(T\to j_{\overline ab}\,j_{\overline cd})$ and $A(T\to j_{\overline ad}\,j_{\overline cb})$, Fourier-transformed vacuum--tetraquark matrix elements of appropriate pairs of quark bilinear currents~(\ref{b}),\begin{equation}\begin{array}{l}\langle0|{\rm T}[j_{\overline ab}(y)\,j_{\overline cd}(y')]|T\rangle\xrightarrow[\text{transformation}]{\text{Fourier}}A(T\to j_{\overline ab}\,j_{\overline cd})\ ,\\[2ex]\langle0|{\rm T}[j_{\overline ad}(y)\,j_{\overline cb}(y')]|T\rangle\xrightarrow[\text{transformation}]{\text{Fourier}}A(T\to j_{\overline ad}\,j_{\overline cb})\ .\end{array}\label{A}\end{equation}\end{itemize}In terms of these hadronic properties, all effective-threshold improved multiquark-adequate QCD sum rules resulting from (once or twice) contracted four-point correlation functions~(\ref{4}) assume, for the example of definitely flavour-exotic tetraquarks, \emph{symbolically} the form~\cite{LMS-A1,LMS-A3} \begin{align}&(f_{\overline ab\overline cd})^2\exp(-M^2\,\tau)\nonumber\\&=\int\limits_{\hat s}^{s_{\rm eff}(\tau)}{\rm d}s\exp(-s\,\tau)\,\rho_{\rm p}(s)+\mbox{Borel-transformed power corrections}\ ,\\&f_{\overline ab\overline cd}\,A(T\to j_{\overline ab}\,j_{\overline cd})\exp(-M^2\,\tau)\nonumber\\&=\int\limits_{\hat s}^{s_{\rm eff}(\tau)}{\rm d}s\exp(-s\,\tau)\,\Delta_{\rm p}(s)+\mbox{Borel-transformed power corrections}\ ,\\&f_{\overline ab\overline cd}\,f_{\overline ad\overline cb}\exp(-M^2\,\tau)\nonumber\\&=\int\limits_{\hat s}^{s_{\rm eff}(\tau)}{\rm d}s\exp(-s\,\tau)\,\rho_{\rm r}(s)+\mbox{Borel-transformed power corrections}\ ,\\&f_{\overline ad\overline cb}\,A(T\to j_{\overline ab}\,j_{\overline cd})\exp(-M^2\,\tau)\nonumber\\&=\int\limits_{\hat s}^{s_{\rm eff}(\tau)}{\rm d}s\exp(-s\,\tau)\,\Delta_{\rm r}(s)+\mbox{Borel-transformed power corrections}\ .\end{align}

The general lesson to be learned from the above for both \emph{perturbative and nonperturbative} QCD contributions to QCD sum-rule approaches applied to \emph{any} type of multiquark hadrons: paying attention to deploy exclusively spectral densities and power corrections computed~in multiquark-phile manner should avoid or, at least, diminish the ``contamination'' of inferred QCD sum-rule predictions by input not related at all to the multiquark hadrons under~study.

\section{Summary, Conclusion and Outlook --- Multiquark-Instigated Theoretical Adaptations}The multiquark states among the conceivable exotic hadrons feature a characteristic not shared by conventional hadrons, namely, cluster reducibility \cite{LMS-C,LMS-A5}, that is to say, their ability to fragment into \emph{colour-singlet} bound states of lesser numbers of constituents, eventually into a set of conventional hadrons. A promising implication for various theoretical approaches to multiquarks is the advantage gained by pertinent modification of one's favoured formalism.

Here, such improvements have been illustrated for the set of flavour-exotic tetraquarks. An analogous contemplation can be (and has been) done for the class of flavour-cryptoexotic tetraquarks \cite{LMS-N1,LMS-N2,LMS-N3,LMS-N3+,LMS-N4,LMS-N7}. It goes without saying that there one gets confronted with additional complications: the potential mixing of these tetraquark states with conventional mesons that carry precisely the quantum numbers of those tetraquarks. Mutatis mutandis, these~findings should be straightforwardly transferable to any other multiquark states, such as the likewise established \cite{PDG} pentaquark baryons. The numerical impact of proposed changes may only be quantified by confronting (definite) multiquark predictions with experimental counterparts.

\vspace{6pt}

\funding{This research received no external funding.}
\dataavailability{Data sharing not applicable.}
\acknowledgments{The author would like to thank both Dmitri I.\ Melikhov and Hagop Sazdjian, for a particularly pleasurable, enjoyable, and inspiring collaboration on various of the topics covered~above.}
\conflictsofinterest{The author declares no conflict of interest.}
\abbreviations{Abbreviations}{The following abbreviations are used in this manuscript:\\

\noindent\begin{tabular}{@{}ll}LHCb&Large Hadron Collider beauty\\OPE&operator product expansion\\QCD&quantum chromodynamics\end{tabular}}

\begin{adjustwidth}{-\extralength}{0cm}

\reftitle{References}

\PublishersNote{}
\end{adjustwidth}
\end{document}